\documentclass[prl,twocolumn]{revtex4-1}
\usepackage{amsmath,amssymb,amsfonts,bm,amscd,color,slashed}
\usepackage{mathtools}
\usepackage{todonotes}
\usepackage{hyperref}
\usepackage{graphics}

\makeatletter
\renewcommand{\@makecaption}[2]{
  \vskip\abovecaptionskip
  \sbox\@tempboxa{\small\sf #1: #2}%
  \ifdim \wd\@tempboxa >\hsize
  \small\sf #1: #2\par
  \else
    \global \@minipagefalse
    \hb@xt@\hsize{\hfil\box\@tempboxa\hfil}%
  \fi
  \vskip\belowcaptionskip}
\makeatother

\def\ba{\begin{eqnarray}}
\def\ea{\end{eqnarray}}

\def\tilde{\widetilde}
\def\hat{\widehat}
\def\bar{\overline}

\def\Dslash{\,\,{\raise.15ex\hbox{/}\mkern-12mu D}}
\def\Dbarslash{\,\,{\raise.15ex\hbox{/}\mkern-12mu {\bar D}}}
\def\delslash{\,\,{\raise.15ex\hbox{/}\mkern-9mu \partial}}
\def\delbarslash{\,\,{\raise.15ex\hbox{/}\mkern-9mu {\bar\partial}}}
\def\pslash{\,\,{\raise.15ex\hbox{/}\mkern-9mu p}}
\def\calDslash{\,\,{\raise.15ex\hbox{/}\mkern-12mu {\cal D}}}

\newcommand{\R}{{\mathbb R}}
\newcommand{\C}{{\mathbb C}}

\def\Tr{{\rm Tr \,}}
\def\dd{{\rm d}}

\def\CA{{\mathcal A}}

\def\CG{{\mathcal G}}

\def\CI{{\mathcal I}}

\def\CK{{\mathcal K}}
\def\CL{{\mathcal L}}

\def\CN{{\mathcal N}}

\def\CW{{\mathcal W}}

\renewcommand{\bar}{\overline}
\renewcommand{\hat}{\widehat}

\begin{document}
\preprint{}

\title{Interval reduction and (super)symmetry}

\author{Mykola Dedushenko}
\affiliation{\it Simons Center for Geometry and Physics,
	Stony Brook University, Stony Brook, NY 11794-3636, USA}

\author{Mikhail Litvinov}
\affiliation{\it Department of Physics and Astronomy, Stony Brook University, Stony Brook, NY 11794-3800, USA}

\begin{abstract}
We study three-dimensional quantum field theories on the interval with symmetry-preserving boundary conditions. The physics and symmetries of the effective 2D theory in the IR are the main subjects of this note. We focus on the (super-)Yang-Mills-Chern-Simons (YM-CS) theories with the Dirichlet boundary conditions on both ends. We warm up with the $\mathcal{N}=0$ and $\mathcal{N}=1$ cases flowing to the bosonic and $\mathcal{N}=(0,1)$ WZW models in 2D. Then we study the 3D $\mathcal{N}=2$ YM-CS on the interval with the $\mathcal{N}=(0,2)$ Dirichlet boundaries. It flows to a non-compact version of the $\mathcal{N}=(0,2)$ WZW. We compute its perturbatively exact two-derivative effective action (i.e., the metric and the B-field), and speculate on the possibility of novel non-perturbative effects. We also construct the 2D Landau-Ginzburg models flowing to the similar sigma models.

\end{abstract}


\maketitle

\newcommand{\be}{\begin{equation}}
\newcommand{\ee}{\end{equation}}

\section{Introduction}
Boundaries and defects form a rich class of observables in quantum field theories (QFT) providing a lot of insight into their dynamics.
They are often associated with symmetries breaking of a theory, which can be explicit and classical, spontaneous, or due to anomalies.
Such phenomena may depend strongly on the spacetime dimensionality, especially as far as spontaneous \cite{Mermin:1966fe} and anomalous breaking \cite{Adler:1969gk,Bell:1969ts,Alvarez-Gaume:1983ihn} are concerned, and defects facilitate interactions across dimensions. They can also mix different symmetry-breaking effects, as in the anomaly inflow mechanism \cite{Callan:1984sa}, where the classical and the anomalous breaking of the same symmetry cancel each other along the boundary or the defect. 
The subject of defect QFT's is vast and goes well beyond the scope of this note. Here we focus on its specific corner, namely, boundaries preserving some fraction of supersymmetry, see \cite{Gaiotto:2008sa,Gadde:2013wq,Okazaki:2013kaa,Yoshida:2014ssa,Dimofte:2017tpi} for a number of examples, including their interplay with global symmetries.

For a local theory on $\R^{D-1}\times \R_+$, the faraway region is described by a $D$-dimensional QFT, whereas close to the boundary we find a hybrid of the $D$ and $(D-1)$-dimensional physics.
If we introduce second parallel boundary, i.e., put our theory on an interval, we obtain another interesting setup, -- that of the interval reduction.
It is perhaps not the first example of dimensional reduction we learn from textbooks. However, examples of QFT on an interval are ubiquitous in physics, both in a lab and on theorists' chalkboards.
An extremely incomplete list of examples includes: Condensed matter experiments, such as those on quantum Hall effect, performed on a slab of material \cite{vonKlitzing:1980pdk,H40yrs}; measurement of the Casimir force between two parallel plates \cite{Sparnaay:1958wg,Bressi:2002fr}, etc. In String Theory, we often construct QFTs via brane engineering \cite{Hanany:1996ie,Witten:1997sc}, and suspending a brane between two other branes is clearly an instance of interval reduction. 
Another example is embedding the heterotic strings into the M-theory, which according to Horava and Witten is done via the interval reduction of the 11D theory \cite{Horava:1995qa}. Topological QFTs (TFTs) on the interval also feature prominently in the recent developments involving symmetry TFTs \cite{Freed:2022qnc,Apruzzi:2021nmk,vanBeest:2022fss}. Our own interest stems from the applications to VOA$[M_4]$ \cite{Dedushenko:2017tdw}, and we explain in more detail in the companion paper \cite{DL2} how the interval reduction can help to compute it for some four-manifolds (see also \cite{Gaiotto:2017euk,Prochazka:2017qum} for related brane setups).

In the UV, the interval-reduced theory looks like a $D$-dimensional/$(D-1)$-dimensional coupled system of the bulk and two boundaries, while at long distances, (in the IR,) it flows to some $(D-1)$-dimensional QFT. This is just like in the usual dimensional reduction, with the exception that signatures of the $(D-1)$-dimensional physics are already present in the UV due to the boundaries. Indeed, if the boundaries support some anomalies in the UV, the 't Hooft anomaly matching condition says that they are robust along the RG flow and match the anomalies of the IR $(D-1)$-dimensional QFT. Such matching provides an important constraint on the IR physics. Another remark we make is that the patterns of symmetry breaking can change as we flow to the $(D-1)$-dimensional theory, and one can even find that the symmetry is restored in the IR.

To be more specific, consider a $D=(2+1)$-dimensional QFT reduced on an interval.
Suppose the bulk theory spontaneously breaks continuous global symmetry $G$ via a vev of some scalar.
Further assume that the theory admits a symmetric boundary condition, meaning that neither explicit nor anomalous breaking of $G$ occurs along the boundary.
What happens to the spontaneous breaking of $G$ in its presence? If there is only one boundary, the breaking still occurs: Far away from the boundary we, as usual, fix the 3d vacuum, which breaks $G$ spontaneously.
Now consider two such parallel boundaries, i.e., put our theory on the interval. Then it becomes macroscopically $(1+1)$-dimensional, and the Coleman-Mermin-Wagner theorem, under the standard assumptions on QFT (such as Wightman axioms), rules out spontaneous symmetry breaking (SSB). Thus in this case, the presence of two symmetric boundaries ``restores'' $G$ that breaks spontaneously in the original 3D theory on $\R^{2,1}$. 
To exemplify consider a complex scalar $\phi$ with the ``hat'' potential $V(\phi) = (|\phi|^2 - v^2)^2$, which of course exhibits SSB of the $U(1)$ global symmetry in 3D. For the interval reduction, choose Neumann boundary conditions on $\phi$, which preserve the $U(1)$. After the reduction we end up with the same scalar field theory in 2D, which, however, cannot break $U(1)$ due to peculiarities of the 2D dynamics. The interval reduction restores the symmetry here.

The answer is not so obvious in the case of supersymmetry, since spontaneous SUSY breaking \cite{Fayet:1974jb,Fayet:1975ki,ORaifeartaigh:1975nky,Witten:1981nf} is possible in 2D \cite{Dine:1986zy,Dine:1987bq}. It is characterized by the positive vacuum energy, and one can easily construct examples that spontaneously break SUSY both in 3D and after the interval reduction to 2D. Indeed, suppose the energy density of the SUSY breaking 3D vacuum is $\rho>0$.
After reduction on the sufficiently large interval of size $L$, the 2D vacuum energy density becomes $L\rho + \rho_\partial$, where $\rho_\partial$ captures the effects of supersymmetric boundary, such as the contribution of boundary degrees of freedom, the Casimir effect, etc.
Importantly, $\rho_\partial$ does not grow with $L$, therefore, for large $L$ the first term dominates and the vacuum energy density remains positive.
Thus reduction on the large interval (equivalently, first flowing to the IR and then reducing on the interval,) results in a SUSY-breaking 2D theory, whenever the original theory breaks SUSY. Whether this conclusion persists for smaller values of $L$ is not obvious, and it might very well happen that at some critical length $L_c$, the theory transitions into the SUSY-preserving phase. What happens in a given 3D theory is thus an interesting dynamical question. For a related discussion in the 4D/3D system see \cite{Igarashi:1984ux}.

Our main focus in the following sections will be a three-dimensional gauge theory with $\CN=2$ supersymmetry reduced on the interval. Namely, consider the pure 3d $\CN=2$ super Yang-Mills with a simple gauge group $G$ and a Chern-Simons level $k$.
It is believed to exhibit the runaway behavior at $k=0$ \cite{Affleck:1982as}, spontaneously break SUSY by the monopole effects for $0<|k|<h$, and preserve SUSY for $|k|\geq h$ \cite{Bergman_1999,Ohta:1999iv}.
We will reduce it on the interval with the $\CN=(0,2)$ Dirichlet boundary conditions imposed on both ends (for Neumann boundaries, see \cite{Sugiyama:2020uqh}).
The effective two-dimensional $\CN=(0,2)$ description, as we will argue, is a non-linear sigma model (NLSM) into the \emph{complexified} group $G_\C$, with the $B$-field given by the Wess-Zumino (WZ) term of level $k$. This is a new \emph{non-compact} version of the $\CN=(0,2)$ WZW \cite{Spindel:1988nh,Spindel:1988sr,Hull:1990qf,Rocek:1991vk,Rocek:1991az}. (For more exotic cases of the WZW-like structures emerging from the higher-dimensional SUSY theories, see \cite{Nekrasov:2021tik,Jeong:2021rll}.) 

The known SUSY-breaking effects (instantons) in 2D $\CN=(0,2)$ NLSMs, --- given by the world sheet wrapping holomorphic curves in the target \cite{Dine:1986zy,Dine:1987bq,Beasley:2005iu}, --- are absent in the $G_\C$ NLSM. This makes the IR behavior in the range $0<|k|<h$ slightly mysterious, as there is a tantalizing possibility that the SUSY is preserved on the interval (even though it is spontaneously broken in 3D). The analysis is also complicated by the fact that the model is noncompact and thus lacks a normalizable vacuum for all values of $k$. We will focus on the range $|k|\geq h$ here, in which the model is expected to preserve SUSY both in 3D and 2D. The case of $|k|<h$ will be addressed elsewhere, but we do believe that the 2D models break SUSY in this range via the novel vortex effects, for which the noncompactness of $G_\C$ is crucial. In short, for some cocharacter $b: \C^\times \to G_\C$, one defines a half-BPS defect by demanding that the NLSM field $\phi$ behaves as
\begin{equation}
\label{vortex_op}
\phi \sim z^b
\end{equation}
near the defect. This makes sense precisely for the complexified $G_\C$ taken as a target, and is the interval-reduced image of the boundary monopole operator in 3D (see also the companion paper \cite{DL2}). It would be interesting to explore such defects.

In the rest of this note, we study the interval reduction of gauge theory quantitatively.
In particular, we compute the (perturbatively exact) two-derivative effective action of the 2D NLSM.
Such a computation is based on a new technical trick that affords a lot of simplifications to our problem. We start with a non-SUSY example, where we explain this trick in detail. It consists of two steps. First go to the gauge in which the gauge field component $A_y$ along the interval vanishes. This necessarily modifies the boundary conditions since the gauge field may have a non-zero Wilson line $g(x^0,x^1)$ along the interval. We would like to integrate out the gauge field at a fixed profile of $g(x)$. In the second step, we notice that positive powers of the length $L$ of the interval only multiply higher-derivative terms in the effective 2D action. Treating $L$ as a parameter in the Lagrangian, we send $L\to 0$ (first subtracting the Casimir energy) to isolate the two-derivative terms. This makes the path integral over gauge fields rather simple, namely, Gaussian. We compute it and find the two-derivative effective action.

This method generalizes straightforwardly to $\CN=1$ and $\CN=2$. The results are most interesting in the $\CN=2$ case, where we compute the nontrivial bi-invariant metric and the B-field on $G_\C$ quite explicitly. We also construct the two-dimensional Landau-Ginzburg (LG) models, which flow to the $\CN=(0,2)$ NLSMs into $G_\C$ as well. We comment on the structure of current multiplets and their relation to the noncompactness of these theories. The LG models seem to be dual to our interval theories perturbatively, however, at the nonperturbative level the duality is not expected. In a sense, they provide a distinct UV definition of the $G_\C$ NLSM, which lacks the vortices \eqref{vortex_op} in the spectrum. A number of conjectures and  speculations are made along the way.

\section{Gauge fields on the interval}
As a warm-up, consider a pure gauge theory on the interval parameterized by $y\in [0,L]$ with Dirichlet boundary conditions on both ends. The dynamical field $g(x)$ of the reduced theory is the holonomy along this interval. Thus the IR theory must be NLSM into the group $G$. Most cleanly this is seen in the 2D Yang-Mills theory:
\begin{equation}
S = \frac1{2e^2} \Tr \int F\wedge * F,
\end{equation}
defined on a strip $\R \times [0,L]$, where $\R$ is the time direction parameterized by $t$. The Dirichlet boundary conditions:
\begin{equation}
A_t\big|=0,
\end{equation}
are imposed both at $y=0$ and $y=L$. Naturally, the group of gauge transformations is
\begin{equation}
\CG = \{g(t,y): \R\times I \to G, g(t,0) = g(t,L)=1\}.
\end{equation}
The holonomy between the two boundaries is
\begin{equation}
g(t) = {\rm P}\exp i\int_L^0 A_y \dd y.
\end{equation}
For a fixed profile $g(t)$, we integrate out $A_\mu$ to find the effective action for $g(t)$. The problem is simplified by the following trick. Let us first perform an ``illegal'' gauge transformation into the gauge $A_y=0$. It is illegal in the sense that it does not belong to the group $\CG$ and cannot preserve both boundary conditions $A_t\big|_{y=0}=0$ and $A_t\big|_{y=L}=0$. Indeed, setting $A_y$ to zero by a gauge transformation amounts to solving $A_y = h^{-1} \partial_y h$, or
\begin{equation}
\label{trick_gauge}
\partial_y h = h A_y.
\end{equation}
If we choose to keep $A_t\big|_{y=0}=0$, then \eqref{trick_gauge} is supplied by the initial condition $h\big|_{y=0}=1$ that yields a Cauchy problem with the unique solution $h=h_\ell(t,y)$ (this also fully fixes gauge).  Likewise preserving $A_t\big|_{y=L}=0$ implies another unique solution $h_r$ with $h_r\big|_{y=L}=1$. At this point, we could subdivide our interval into patches $(0,a] \cup [a, L)$, perform the gauge transformation $h_\ell$ on $(0,a]$, and $h_r$ -- along $[a,L)$. This will set $A_y=0$ everywhere, preserve the boundary conditions $A_t\big|=0$ on both ends. However, at $y=a$ the two patches are glued by the gauge transformation $g(t)$. The latter is obvious because the Wilson line connecting the two boundaries is gauge-invariant (and must be equal to $g(t)$).

The location $y=a$ of the gluing surface is arbitrary, and we can collide it with $y=L$, i.e. send $a \to L$. This is equivalent to simply performing the gauge transformation $h_\ell$ everywhere on $(0,L)$.
As a result, we obtain a single gauge patch with the modified boundary conditions at $y=L$  determined by $g(t)$:
\begin{equation}
A_t\big|_{y=0} = 0,\quad A_t\big|_{y=L}=g^{-1}(t) \partial_t g(t).
\end{equation} 
With such boundary conditions and in the gauge $A_y=0$, there is no remaining gauge freedom. The Yang-Mills action becomes quadratic:
\begin{equation}
S = \frac1{2e^2} \int \Tr (\partial_y A_t)^2 \dd y\,\dd t.
\end{equation}
Thus $A_t$ is easily integrated out by solving the equations of motion (EOM), while the determinant is a constant that can be dropped (which in fact cancels against the constant Faddeev-Popov determinant associated with $A_y=0$).
The classical solution is
\begin{equation}
A_t = \frac{y}{L}  g^{-1}(t) \partial_t g(t),
\end{equation}
resulting in the 1D action:
\begin{equation}
\label{eff_1d}
S_{\rm 1D} = \frac1{2e^2 L}\int \Tr (g^{-1} \partial_t g)^2 \dd t.
\end{equation}
This computation is, of course, completely exact, and the answer \eqref{eff_1d} is expected based on the $G\times G$ global symmetry of the interval theory.

In the 3D Yang-Mills case, the analysis is quite similar, except it cannot be performed exactly. The 2D effective action can be computed in the two-derivative approximation, ignoring the higher-derivative corrections. We work on $\R^2 \times I$ with coordinates $(x^0, x^1, y)$ and with the boundary conditions $A_0\big| = A_1\big|=0$. Denoting the holonomy along $I$ by $g(x)$ and passing to the gauge $A_y=0$, we have the analogous boundary conditions:
\begin{equation}
\label{bc0g}
A_i \big|_{y=0} = 0,\quad A_i \big|_{y=L} = g^{-1} \partial_i g,\quad i=0,1.
\end{equation}
After rescaling the interval coordinate as $y=L\xi$, the YM action takes the form:
\begin{equation}
\frac1{2e^2 L} \Tr \int \dd^2 x\int_0^1 \dd\xi \left[ (\partial_\xi A_i)^2 + L^2(D_0 A_1 - \partial_1 A_0)^2 \right].
\end{equation}
We regard $e^2 L = \lambda^2$ as a dimensionless coupling, while $L$ is a dimension-length parameter. Since $g(x)$ is dimensionless, the expansion in powers of $L$ is the derivative expansion of the 2D effective action. There is a caveat: If $\Lambda$ is a UV momentum cut-off, the powers of $L$ could be also compensated by $\Lambda$, giving another dimensionless parameter $L\Lambda$. The positive powers of $\Lambda$ would signal the power-law UV divergences, normally canceled by the UV counterterms. However, the 3d Yang-Mills is UV finite and has no counterterms \cite{Dudal:2004ch,Dudal:2006ip,Collins:1974bg}. Thus the power-law UV divergences simply cannot appear in the effective action, so the positive powers of $L\Lambda$ are absent, and the expansion in powers of $L$ is indeed the derivative expansion. The possible $O(L^{-2})$ Casimir term should be subtracted by hand. Then at the leading $O(L^0)$ order we find the two-derivative effective action, so we drop the irrelevant higher-derivative terms simply by setting $L=0$. At this order, the microscopic action becomes $\frac1{2\lambda^2}\int\dd^2 x\, \dd\xi (\partial_\xi A_i)^2$, which has the saddle point $A_i = \xi g^{-1}\partial_i g$, so we obtain:
\begin{equation}
S_0^{\rm eff} = \frac1{2 \lambda^2} \int \dd^2 x\, \Tr(g^{-1} \partial_i g)^2.
\end{equation}
Thus, the two-derivative effective action is exactly captured by the \emph{principal chiral model} (PCM).

Let us upgrade this analysis to also include the Chern-Simons (CS) term:
\begin{equation}
S_{\rm CS} = \frac{k}{4\pi} \Tr \int \left[ A\dd A + \frac23 A^3 \right].
\end{equation}
Starting with the same boundary conditions and passing to the gauge $A_y=0$, we again have \eqref{bc0g}. The CS term becomes $\frac{k}{4\pi} \Tr \int A \dd_y A + k S_{\rm WZ}[g]$, where $\dd_y \equiv \dd y \frac{\partial}{\partial y}$ and $S_{\rm WZ}[g]$ is the 2D Wess-Zumino (WZ) term generated by the gauge transformation of the CS action in the presence of the boundary. The full action then becomes:
\begin{equation}
\begin{split}
\hspace*{-0.2cm} S &= \frac{1}{2\lambda^2} \Tr\int \dd^2x\,\dd\xi \Big[ (\partial_\xi A_i)^2 + L^2(D_0 A_1 - \partial_1 A_0)^2\\ &+  \frac{k\lambda^2}{2\pi}(A_1\partial_\xi A_0 - A_0\partial_\xi A_1)  \Big] + k S_{\rm WZ}[g].
\end{split}
\end{equation}
Again the term $L^2 F_{01}^2$ is dropped in the two-derivative approximation. The saddle point equations in Minkowski signature become:
\begin{equation}
\partial_\xi^2 A_1 = -2\omega\partial_\xi A_0,\quad \partial_\xi^2 A_0 = -2\omega\partial_\xi A_1,
\end{equation}
where
\begin{equation}
\omega = \frac{k e^2 L}{4\pi}\equiv \frac{k\lambda^2}{4\pi}.
\end{equation}
Subject to the same boundary conditions \eqref{bc0g}, these equations are easily solved, leading to the 2D action:
\begin{equation}
\label{WZWbos}
S_0^{\rm eff} = \frac{k}{8\pi} \frac{1}{\tanh\omega} \int \dd^2x\, \Tr (g^{-1}\partial_i g)^2 + k S_{\rm WZ}[g].
\end{equation}
Again, the global $G\times G$ symmetry of the interval theory was obvious from the beginning. Since a simple compact Lie group $G$ has a unique, up to an overall scale, bi-invariant metric $\Tr (g^{-1} \dd g)^2$, the kinetic term in $S_0^{\rm eff}$ was bound to take this form. Only the coefficient in front of the action is a nontrivial result of our computation.

What we obtained is, naturally, the WZW$_k$ model, which then flows to the conformal point \cite{Witten:1983ar}.
Of course we expected this.
Starting with a long interval, we could first flow to the IR in 3D by dropping the irrelevant YM term, ending up with the CS on an interval, which is known to yield the WZW$_k$ \cite{Elitzur:1989nr}.

\section{Minimal supersymmetry}
Before studying $\CN=2$, let us briefly look at the $\CN=1$ SYM at level $k$ also considered in \cite{Gaiotto:2019asa}. The Lagrangian consists of the same bosonic part as above, plus the action for massive adjoint Majorana gaugini $\chi$:
\begin{equation}
\CL_f = i \Tr \chi \slashed{D}\chi - \frac{ke^2}{2\pi} \Tr \chi\chi.
\end{equation}
Again we pass to the gauge $A_y=0$ and rescale $y=L\xi$, after which the fermion action becomes
\begin{equation*}
  S_f = \frac1{2e^2}\int\dd^2 x\, \dd\xi\, \Tr \left[ i\chi \gamma^2 \partial_\xi \chi - \frac{k \lambda^2}{2\pi} \chi\chi + iL \chi\gamma^i D_i \chi \right]\!{.}
\end{equation*}
We impose the $\CN=(0,1)$ version of the Dirichlet boundary conditions, and the interval zero mode of $\chi$ is found by solving
\begin{equation}
\label{chi_zm}
\partial_\xi \chi + \frac{k\lambda^2}{2\pi} i\gamma^2 \chi=0.
\end{equation}
It is identified with a chiral (right-moving) edge mode living either on the left or on the right boundary, depending on the sign of $k$. The interval non-zero modes look from the 2D perspective like heavy fermions (with masses of order $1/L$) that can only contribute higher-derivative terms in the effective action. We thus conclude that the 2-derivative 2D effective action \eqref{WZWbos} is only slightly modified by a chiral Fermi kinetic term (plus a superpartner of the WZ term \cite{Abdalla:1984ef,DiVecchia:1984nyg,Spindel:1988nh,Spindel:1988sr,Hull:1990qf,Rocek:1991vk,Rocek:1991az,Gaiotto:2019asa}).

Naturally, the 2D limit is expected to be the $\CN=(0,1)$ WZW$_k$, and precisely such an interval reduction was also considered in \cite{Gaiotto:2019asa}. The 't Hooft anomaly for the boundary $G\times G$ global symmetry is
\begin{equation}
\label{anom01}
P = \left(k - \frac{h^\vee}{2}\right)\Tr F_\ell^2 - \left(k + \frac{h^\vee}{2}\right)\Tr F_r^2,
\end{equation}
where the $k$ contribution is from the inflow and the $\frac{h^\vee}{2}$ is from the boundary anomalies of fermions. Here $F_\ell$ and $F_r$ are the curvatures associated to global symmetries on the left and right boundaries. The $\frac{h^\vee}{2}$ contribution to the anomaly is clearly matched by the 2D Majorana-Weyl fermions in the $(0,1)$ $G$-valued scalar multiplets. The $k$ contribution is matched by the 2D WZ term. Such anomaly matching agrees with the $\CN=(0,1)$ WZW$_k$ proposal. As was noted in \cite{Gaiotto:2019asa}, regardless of the sign of the anomalies, the affine current algebras one finds in the IR must have positive levels. This is because the model is \emph{compact} (as $G$ is compact), so the standard unitarity constraints apply. In particular, for $k\geq \frac{h^\vee}{2}$ one finds the affine symmetry $\hat{\mathfrak{g}}_{k-h^\vee/2} \oplus \hat{\mathfrak{g}}_{k+h^\vee/2}$ in the left-moving and right-moving sectors at the CFT point, respectively. For $k\leq -\frac{h^\vee}{2}$, one finds $\hat{\mathfrak{g}}_{-k-h^\vee/2} \oplus \hat{\mathfrak{g}}_{-k+h^\vee/2}$ in the left and right sectors, respectively. Note that $k\pm \frac{h^\vee}{2}$ are assumed to be integers, by the usual parity anomaly considerations in 3D. When $|k|<\frac{h^\vee}{2}$, the dynamical SUSY breaking is expected in two dimensions \cite{Gaiotto:2019asa}.

Notice that here, the fate of SUSY in the 2D model mimics what happens in the parent 3D $\CN=1$ theory, which also breaks it for $|k|< \frac{h^\vee}{2}$ \cite{Witten:1999ds}.
For $|k|\geq\frac{h^\vee}{2}$, the 3D theory flows to the level-$\left(k-\frac{h^\vee}{2}{\rm sgn}(k)\right)$ CS (and is trivially gapped for $|k|=\frac{h^\vee}{2}$) \cite{Witten:1999ds,Gomis:2017ixy}.
At each end of the interval, one then naturally finds bosonic level-$\left|k-\frac{h^\vee}{2}{\rm sgn}(k)\right|$ WZW currents.
Additionally, one (and only one) of the two boundaries contains a set of chiral $\mathbf{adj}(G)$-valued fermionic edge modes, as determined by \eqref{chi_zm}.
 They supersymmetrize the WZW currents on that boundary.
 This clearly matches the IR physics of the $\CN=(0,1)$ NLSM into $G$ with the level-$k$ WZ term (the $h^\vee/2$ shift comes from the fermions).
 The 3D IR physics for $|k|<\frac{h^\vee}{2}$ is also known to be described by a certain TQFT plus a decoupled subsector of Goldstino modes \cite{Gomis:2017ixy,Bashmakov:2018wts}.
 The 2D limit in this case is some CFT (as evidenced by the anomaly \eqref{anom01} for continuous symmetries), which is non-SUSY and is not studied here.

 \section{\texorpdfstring{$\CN=2$}{N=2} case}
Our main subject is the $\CN=2$ YM-CS with gauge group $G$ at level $k$, with $\CN=(0,2)$ Dirichlet boundary conditions imposed on the vector multiplet at both ends of the interval (See \cite{Armoni:2015jsa} for brane realizations of such setups). Thinking of the 3D vector as $(V,S)$, where $V$ is a 2D $(0,2)$ vector and $S$ is a 2D $(0,2)$ chiral \cite{Dimofte:2017tpi}, both valued in the gauge group ${\rm Hom}(\R, G)$, the boundary conditions eliminate $V$ along the boundary. The lowest component of $S$ is
\begin{equation}
\CA_y = A_y + i\sigma,
\end{equation}
where $\sigma$ is the real scalar in the 3D $\CN=2$ vector multiplet. Fields that remain dynamical in the IR limit are the interval zero modes of $S$. In particular, the natural gauge-invariant bosonic variable is the complexified open Wilson line:
\begin{equation}
g(x) = {\rm  P}\exp i\int_L^0 \CA_y \dd y.
\end{equation}
The 2D $\CN=(0,2)$ SUSY completes this into the chiral multiplet (roughly given by ${\rm P}\exp \int_L^0 S \dd y$). Thus in the IR, we expect to find the 2D $(0,2)$ NLSM into $G_\C$ -- the complexification of $G$ whose Lie algebra is $\mathfrak{g}_\C=\mathfrak{g}\otimes\C$. Furthermore, the 2D action must include a B-field given by the WZ term at level $k$ (plus superpartners), similar to the bosonic case \eqref{WZWbos}.

This proposal passes simple checks via anomalies. The anomaly of global $G\times G$ symmetry can be computed from the UV gauge theory description as:
\begin{equation}
\label{anom02}
P = -(k+h^\vee)\Tr F_\ell^2 + (k-h^\vee) \Tr F_r^2,
\end{equation}
where $F_\ell$ and $F_r$ are the curvatures associated to global symmetries of the left and right boundaries.\! In the NLSM description, $(g_\ell, g_r) \in G\times G$ acts according to:
\begin{equation}
(g_\ell, g_r): g(x) \mapsto g_\ell g(x) g_r^{-1},
\end{equation}
and to compute its perturbative anomaly, it is enough to work locally on the target $G_\C$, where the $\mathfrak{g}\oplus \mathfrak{g}$ symmetry is represented by the Killing vector fields. The contribution $-h^\vee(\Tr F_\ell^2 + \Tr F_r^2)$ comes from the right-moving fermions, and $k(\Tr F_r^2 - \Tr F_\ell^2)$ is saturated by the WZ term, confirming that the anomalies match.

The theory on $\R^{2,1}$ is fully gapped: All particles, including the photon, have masses proportional to the CS level.
On the interval, the gap closes, as evidenced by the boundary 't Hooft anomalies $P$ computed above.
The interval zero modes in $S$ can be identified with the edge modes living on one of the two boundaries of the interval (depending on the sign of $k$).
Let us show how the zero mode of $\sigma$ comes about. In the presence of CS level, the boundary condition on $\sigma$ is of Robin type \cite{Dimofte:2017tpi,RobinBC:2017}:
\begin{equation}
D_y\sigma\big| = \frac{ke^2}{2\pi}\sigma\big|.
\end{equation}
Ignoring fermions, its equation of motion (EOM) is
\begin{equation}
D^\mu D_\mu \sigma + m^2\sigma = 0,
\end{equation}
where $m = \frac{ke^2}{2\pi}$. If $k<0$, we find an edge mode supported at $y=0$:
\begin{equation}
\sigma = \sigma_{\rm edge}(x) e^{my},
\end{equation}
and if $k>0$, it is supported at $y=L$:
\begin{equation}
\sigma = \sigma_{\rm edge}(x) e^{-m(L-y)}.
\end{equation}
In either case, the EOM reduces to
\begin{equation}
D^i D_i \sigma_{\rm edge}(x)=0,
\end{equation}
showing that $\sigma_{\rm edge}$ is indeed the massless edge mode of $\sigma$. Together with the zero mode of $A_y$ and fermions, they describe the $\CN=(0,2)$ chiral edge modes (or interval zero modes) in $S$, valued in the complexified gauge group $G_\C$, making the IR NLSM non-compact.

Note that this clarifies the somewhat subtle $(0,2)$ Dirichlet boundary with the ``wrong'' sign of the CS level, previously encountered in \cite{Costello:2020ndc}. Such a boundary simply supports a noncompact $G_\C$-valued chiral edge mode $S$. Furthermore, there is an obvious dual description of such boundary conditions: We could consider $\CN=(0,2)$ Neumann boundary conditions instead, coupled to the $G_\C$-valued chiral multiplet along the boundary. The bulk gauge field gauges, say, the right $G$-action on $G_\C$, while the left $G$-action matches the boundary $G$ symmetry of the Dirichlet boundary conditions.

For completeness, we also describe chiral zero modes responsible for the boundary anomalies of the bosonic Maxwell+CS theory.
Consider the EOM for the YM-CS:
\begin{equation}
  \begin{split}
    \dd *F = \left(  \frac{k e^2}{2\pi} \right) F.
  \end{split}
\end{equation}
Using the $ A_{i}| = 0 $ boundary conditions and setting $ A_{i} $ to zero everywhere, we get the following equations:
\begin{equation}
  \begin{split}
    \partial^{ i } F_{iy} = \partial ^{ i }\partial _{i}A_{y} =  0,\\
    \partial _{y}F_{yx_0}  = {m} F_{x_1 y},\\
    \partial _{y}F_{y x_1} = m F_{x_0 y},
  \end{split}
\end{equation}
where $ m=\frac{ke^2}{2\pi} $ is a mass of the gauge boson.
One can rewrite them using light-cone coordinates $ x^{ \pm  } = x_1 \pm x_0 $:
\begin{equation}
  \begin{split}
    \partial _{+}\partial _{-} A_{y} &= 0, \\
    \partial_y \partial _{+} A_{y} &= -m \partial _{+} A_{y},\\
    \partial_y \partial _{-}A_{y} &= m \partial_{-}A_{y}.
  \end{split}
\end{equation}
Thus, we can see from the first equation that $ A_{y} = f( x_+, y) + g (x_-, y) $.
So, the connection zero mode splits into chiral and anti-chiral parts.
Let us solve the equation for the chiral part:
\begin{equation}
  \begin{split}
    \partial _{y} \partial_- g = m \partial_- g.
  \end{split}
\end{equation}
This has the decaying solution that is written either as $ g= e^{ m y }g_0(x_-) $ for $m<0$, or $ g= e^{- m(L- y) }g_0(x_-) $ for $m>0$. 
We thus learn that $ g_0(x_-) $ is a boundary chiral mode. The 't Hooft anomaly for the boundary symmetry is extracted from the OPE of the current $J= \partial_- g_0$ with itself:
\begin{equation}
  \begin{split}
    J \left(z\right)J \left(w\right) \sim \frac{k}{(z-w)^{2 }}.
  \end{split}
\end{equation}
That gives rise to the usual CS anomaly $ k $ discussed earlier in this section.
By the similar argument, the anti-chiral mode $f_0(x_+)$ lives on the opposite boundary and gives the opposite $ -k $ contribution to the anomaly.
\subsection{Effective 2D action}
We can repeat our exercise and derive the effective $\CN=(0,2)$ NLSM into $G_\C$ by integrating out the gauge multiplet on the interval. This problem becomes significantly more cumbersome in the $\CN=2$ case, but luckily, it is still solvable. Another good news is that the Casimir energy term $O(L^{-2})$ is absent due to SUSY \cite{Igarashi:1984ux}. We consider the $\CN=2$ YM-CS action and integrate out the auxiliary field $D$, which generates the mass term for $\sigma$, so the total action is
\begin{equation}
\begin{split}
\hspace*{-0.2cm}S_{\rm YM} &= \frac1{2e^2}\Tr \int\dd^3 x\bigg[\frac12 F_{\mu\nu}^2 + (D_\mu\sigma)^2 - \frac{(k e^2)^2}{(2\pi)^2} \sigma^2\cr 
&+ i\bar\chi \gamma^\mu D_\mu \chi - \frac{ke^2}{2\pi}\bar\chi\chi + i\chi[\bar\chi,\sigma]\bigg]\cr 
&+ \frac{k}{4\pi} \Tr \int \left(A\dd A + \frac23 A^3\right),
\end{split}
\end{equation}
where the gaugini $\chi$ are Dirac spinors now. Let us perform our favorite trick of passing to the gauge $A_y=0$. Before doing so, we ought to define the complexified and real Wilson lines, respectively:
\begin{equation}
\hspace*{-0.2cm}g(x) = {\rm  P}\exp i\int_L^0 \CA_y \dd y,\quad h(x) = {\rm  P}\exp i\int_L^0 A_y \dd y.
\end{equation}
After the gauge transformation, $A_y$ vanishes and the real (i.e., compact-valued) Wilson line becomes trivial:
\begin{equation}
h(x) \mapsto h(x)^{-1} h(x) e = e,
\end{equation}
where $e\in G$ is the identity element. The way Wilson line transforms reflects the fact that the gauge transformation is trivial at $y=0$ and non-trivial (and equal to $h(x)^{-1}$) at $y=L$. The complexified Wilson line transforms in the same way:
\begin{equation}
  \begin{split}
g \mapsto h^{-1} g e = {\rm P}\exp i\int_0^L (A_y^h + i\sigma^h)\dd y =\\
={\rm P} \exp \int_0^L (-\sigma^h)\dd y,
  \end{split}
\end{equation}
where $\sigma^h = h^{-1} \sigma h$, and $A_y^h = 0$ by definition of $h$. If we define
\begin{equation}
g_\sigma = {\rm P} e^{-\int_0^L \sigma^h \dd y},
\end{equation}
then the above equations imply:
\begin{equation}
g = h g_\sigma.
\end{equation}
In this way, we arrive at the well-known polar, or Cartan, decomposition corresponding to writing the complexified group as
\begin{equation}
G_\C = G \cdot \exp i\mathfrak{g}.
\end{equation}
It is convenient to rescale the fields according to
\begin{equation}
\phi = L \sigma^h,\quad \psi = L\chi^h,\quad \bar\psi = L \bar{\chi}^h,
\end{equation}
where the superscript $h$ denotes conjugation by $h$. We define $y=L\xi$ as before, so that
\begin{equation}
g_\sigma = {\rm P} e^{-\int_0^1 \phi \dd \xi}.
\end{equation}
The global symmetries associated to the two boundaries act on $g$ according to:
\begin{equation}
G\times G\ni(g_\ell, g_r): g \mapsto g_\ell g g_r^{-1}.
\end{equation}
This can be written in terms of $h$ and $\phi$ as follows:
\begin{align}
h \mapsto g_\ell h g_r^{-1},\qquad \phi \mapsto g_r \phi g_r^{-1}.
\end{align}
Note that $\phi$ only transforms under the $g_r$.

It is again convenient to define the 2d coupling:
\begin{equation}
\lambda^2 = e^2 L.
\end{equation}
The action now takes the following form:
\begin{widetext}
\begin{align}
S &= \frac1{2\lambda^2} \Tr \int\dd^2 x\,\dd\xi\bigg[(\partial_\xi A_i)^2 + \frac{k\lambda^2}{2\pi}(A_1\partial_\xi A_0 - A_0\partial_\xi A_1)
+L^2(D_0 A_1 - \partial_1 A_0)^2-\frac1{L^2}\left((\partial_\xi\phi)^2 + \left(\frac{k\lambda^2}{2\pi}\right)^2\phi^2\right)\cr 
&+ \frac1{L}\left(i\bar\psi\gamma^3\partial_\xi \psi + i\psi[\bar\psi,\phi] - \frac{k\lambda^2}{2\pi} \bar\psi\psi\right)
+ (D_i\phi)^2 + i\bar\psi\gamma^i D_i\psi\bigg] + k S_{\rm WZ}[g].
\end{align}
\end{widetext}
The strategy is as before: Perform the saddle point analysis in the $L\to0$ limit and focus on the leading contribution, as the $O(L)$ corrections only produce higher-derivative terms in the effective action. The terms multiplied by the inverse powers of $L$ must vanish in the leading order, determining the interval zero modes for $\phi$ and $\psi,\bar\psi$. 
For simplicity, let us focus on the bosonic part of the action, as the fermionic completion follows by SUSY. Then, in the leading order, we have the path integral with 
\begin{equation}
\begin{split}
S &= \frac1{2\lambda^2} \Tr \int\dd^2 x\,\dd\xi\bigg[(\partial_\xi A_i)^2 + (D_i\phi)^2\\ &+ \frac{k\lambda^2}{2\pi}(A_i\partial_\xi A_0 - A_0\partial_\xi A_1)\bigg] + k S_{\rm WZ}[g],
\end{split}
\end{equation}
subject to the zero mode condition:
\begin{equation}
\partial_\xi^2\phi = \left(\frac{k\lambda^2}{2\pi}\right)^2\phi,
\end{equation}
and the boundary conditions:
\begin{equation}
  \begin{split}
    \partial_\xi \phi\big|_{\xi=0,1} =& \frac{k\lambda^2}{2\pi}\phi\big|_{\xi=0,1},\qquad A_i\big|_{\xi=0}=0\\
    &A_i\big|_{\xi=1}=g^{-1}\partial_i g.
  \end{split}
\end{equation}
The equation on $\phi$ zero mode is solved by:
\begin{equation}
\phi = \varphi(x) e^{\frac{k\lambda^2}{2\pi}\xi},
\end{equation}
where $\varphi(x)$ is a real scalar in the 2d effective action. We do not integrate over $\varphi(x)$ and simply hold it as a degree of freedom in the effective 2d theory. On the contrary, we integrate over the gauge field with the quadratic action:
\begin{equation}
\begin{split}
S_0 &= \frac1{2\lambda^2} \Tr \int\dd^2 x\,\dd\xi \bigg[(\partial_\xi A_i)^2 + 2i\partial_i\phi [A_i,\phi]\\ &- ([A_i,\phi])^2 + \frac{k\lambda^2}{2\pi}(A_1\partial_\xi A_0 - A_0\partial_\xi A_1)
\bigg].
\end{split}
\end{equation}
By varying $A_i$ in this action, we obtain the saddle point equations:
\begin{align}
\label{saddleNeq2}
\partial_\xi^2 A_1 &= -\frac{k\lambda^2}{2\pi}\partial_\xi A_0 + e^{\frac{k\lambda^2}{\pi}\xi}[\varphi,[\varphi,A_1]] + i e^{\frac{k\lambda^2}{\pi}\xi} [\varphi, \partial_1\varphi],\cr \partial_\xi^2 A_0 &= -\frac{k\lambda^2}{2\pi}\partial_\xi A_1 + e^{\frac{k\lambda^2}{\pi}\xi}[\varphi,[\varphi,A_0]] + i e^{\frac{k\lambda^2}{\pi}\xi} [\varphi, \partial_0\varphi].
\end{align}
Let us introduce a notation
\begin{equation}
\hat\varphi = {\rm Ad}_{\varphi}.
\end{equation}
In other words, $\hat\varphi$ is a $(\dim G)\times (\dim G)$ matrix in the adjoint representation of $\mathfrak{g}$, while $\varphi$ and $A_i$ are $(\dim G)$-component vectors representing elements of $\mathfrak{g}$.
 For a moment, let us assume that $\hat\varphi$ is invertible -- this will allow us to solve \eqref{saddleNeq2} explicitly, and the answer will admit an obvious extension to the degenerate $\hat\varphi$.
First we perform a simple shift:
\begin{align}
  A_1 &= a_1 - i \hat\varphi^{-1} \partial_1 \varphi,\\
  A_0 &= a_0 - i \hat\varphi^{-1} \partial_0 \varphi,
\end{align}
which removes the inhomogeneity:
\begin{align*}
\partial_\xi^2 a_1 + \frac{k\lambda^2}{2\pi} \partial_\xi a_0 - e^{\frac{k\lambda^2}{\pi}\xi} \hat\varphi^2 a_1 &= 0,\cr
\partial_\xi^2 a_0 + \frac{k\lambda^2}{2\pi} \partial_\xi a_1 - e^{\frac{k\lambda^2}{\pi}\xi} \hat\varphi^2 a_0 &= 0.
\end{align*}
Defining $a_\pm = a_1 \pm a_0$, we obtain equations:
\begin{equation}
\begin{split}
\partial_\xi^2 a_+ + \frac{k\lambda^2}{2\pi} \partial_\xi a_+ - e^{\frac{k\lambda^2}{\pi}\xi} \hat\varphi^2 a_+ &= 0,\cr
\partial_\xi^2 a_- - \frac{k\lambda^2}{2\pi} \partial_\xi a_- - e^{\frac{k\lambda^2}{\pi}\xi} \hat\varphi^2 a_- &= 0,
\end{split}
\end{equation}
which are easy to solve.
Subject to our boundary conditions, we find the solutions for $A_\pm = A_1\pm A_0$:
\begin{widetext}
\begin{align}
A_- &= \frac{\sinh \left[\frac{2\pi}{k\lambda^2}\left(e^{\frac{k\lambda^2}{2\pi}\xi}-1\right)\widehat{\varphi}\right]}{\sinh \left[\frac{2\pi}{k\lambda^2}\left(e^{\frac{k\lambda^2}{2\pi}}-1\right)\widehat{\varphi}\right]}\left(g^{-1}\partial_- g -
\frac{\cosh \left[\frac{2\pi}{k\lambda^2}\left(e^{\frac{k\lambda^2}{2\pi}}-1\right)\widehat{\varphi}\right]-1}{\widehat{\varphi}}i\partial_-\varphi\right)\cr
&+ \frac{\cosh \left[\frac{2\pi}{k\lambda^2}\left(e^{\frac{k\lambda^2}{2\pi}\xi}-1\right)\widehat{\varphi}\right]-1}{\widehat{\varphi}} i \partial_-\varphi,\\
A_+ &= \frac{\sinh \left[\frac{2\pi}{k\lambda^2}\left(e^{\frac{k\lambda^2}{2\pi}\xi}-1\right)\widehat{\varphi}\right]}{\sinh \left[\frac{2\pi}{k\lambda^2}\left(e^{\frac{k\lambda^2}{2\pi}}-1\right)\widehat{\varphi}\right]}e^{\frac{k\lambda^2}{2\pi}(1-\xi)}\left(g^{-1}\partial_+ g - \frac{\cosh \left[\frac{2\pi}{k\lambda^2}\left(e^{\frac{k\lambda^2}{2\pi}}-1\right)\widehat{\varphi}\right]e^{-\frac{k\lambda^2}{2\pi}}-1}{\widehat{\varphi}}i\partial_+\varphi\right)\cr
&+ \frac{\cosh \left[\frac{2\pi}{k\lambda^2}\left(e^{\frac{k\lambda^2}{2\pi}\xi}-1\right)\widehat{\varphi}\right]e^{-\frac{k\lambda^2}{2\pi}\xi}-1}{\widehat{\varphi}} i \partial_+\varphi.
\end{align}
This solution has a manifestly well-defined $ \hat{\varphi} \to  0 $ limit. The action evaluates to $\frac1{2\lambda^2}\int \dd^2 x\, L + k S_{\rm WZ}[g]$, where
\begin{equation}
L = -\frac12 \Tr (A_+ \partial_\xi A_- + A_- \partial_\xi A_+)\big|_{\xi=1} + \frac{\pi}{k\lambda^2} \left(e^{\frac{k\lambda^2}{\pi}}-1\right)\Tr(\partial_i\varphi)^2 - \frac{i}{2}\Tr\int_0^1 \dd\xi \left\{ A_+ [\varphi, \partial_-\varphi] + A_- [\varphi, \partial_+\varphi] \right\}e^{\frac{k\lambda^2}{\pi}\xi},
\end{equation}
\end{widetext}
which after tedious calculations can be put into the following form:
\begin{equation}
\begin{split}
  \Tr\!&\left(g^{-1} \partial_\nu g \widehat{G}_0 g^{-1}\partial_\mu g + \partial_\nu\varphi \widehat{G}_1 \partial_\mu\varphi +\right. \\
&\left. +  g^{-1}\partial_\nu g \widehat{G}_2 \partial_\mu\varphi \right)\eta^{\mu\nu}
+ \Tr (g^{-1}\partial_\nu g \widehat{B}_1 \partial_\mu\varphi)\varepsilon^{\mu\nu},
\end{split}
\end{equation}
where $\eta^{00}=-\eta^{11}=1$ is the 2D Minkowski metric, $\varepsilon^{01}=-\varepsilon^{10}=1$, and $\widehat{G}_0, \widehat{G}_1, \widehat{G}_2, \widehat{B}_1$ are the $(\dim G)\times (\dim G)$ matrices written in terms of $\widehat{\varphi}$ as follows:
\begin{equation}
\begin{split}
\widehat{G}_0&=\frac{e^{\frac{k\lambda^2}{2\pi}} \widehat{\varphi}}{\tanh\left[\frac{2\pi}{k\lambda^2}\left(e^{\frac{k\lambda^2}{2\pi}}-1\right)\widehat{\varphi}\right]} - \frac{k\lambda^2}{4\pi},\\
\widehat{G}_1 &= \left(e^{\frac{k\lambda^2}{2\pi}}+1\right)\frac{\tanh\left[\frac{\pi}{k\lambda^2}\left(e^{\frac{k\lambda^2}{2\pi}}-1\right)\widehat{\varphi}\right]}{\widehat{\varphi}},\\
\widehat{G}_2 &= - \frac{i k \lambda^2}{2\pi\widehat{\varphi}}\left(1 - \frac{\frac{4\pi \widehat{\varphi}}{k\lambda^2} e^{\frac{k\lambda^2}{2\pi}}}{\tanh\left[\frac{2\pi}{k\lambda^2}\left(e^{\frac{k\lambda^2}{2\pi}}-1\right)\widehat{\varphi}\right]}\right. +  \\
              &\left. +\frac{\frac{2\pi}{k\lambda^2}\left(e^{\frac{k\lambda^2}{2\pi}}+1\right)\widehat{\varphi}}{\sinh\left[\frac{2\pi}{k\lambda^2}\left(e^{\frac{k\lambda^2}{2\pi}}-1\right)\widehat{\varphi}\right]}\right),\cr
\widehat{B}_1 &= \frac{ik\lambda^2}{2\pi\widehat{\varphi}}\left(1 - \frac{\frac{2\pi}{k\lambda^2}(e^{\frac{k\lambda^2}{2\pi}}-1)\widehat{\varphi}}{\sinh\left[\frac{2\pi}{k\lambda^2}(e^{\frac{k\lambda^2}{2\pi}}-1)\widehat{\varphi}\right]}\right).
\end{split}
\end{equation}
For example, the first term, with $(g^{-1} \dd g)^a\equiv \theta^a$ being the left-invariant one-forms, can be written as $\theta^a_\nu (\widehat{G}_0)_{ab} \theta^b_\mu \eta^{\mu\nu}$, where $a,b=1..\dim G$.
Also note that the matrix elements of $\widehat{\varphi}$ are $\widehat{\varphi}_a{}^b = \sum_{c=1}^{\dim G} \varphi^c\, i f_{ca}{}^b$, that is $\widehat{\varphi}$ is a purely imaginary antisymmetric matrix.

To summarize, the effective action we have found describes the $\CN=(0,2)$ WZW model into $G_\C$, with the particular choice of bi-invariant metric and the B-field given above. The curvature of the B-field is cohomologous to the usual three-form $k f_{abc}\theta^a\wedge\theta^b\wedge\theta^c$, which is pulled back from the compact subgroup $G\subset G_\C$. Note that this WZW model is different from the $\CN=(0,2)$ WZW models known in the literature \cite{Spindel:1988nh,Spindel:1988sr,Hull:1990qf,Rocek:1991vk,Rocek:1991az} (see also \cite{Faizal:2016skd}), which are sigma models into the compact groups admitting complex structure.

\subsection{Non-compactness}
The main distinction, both from the known $\CN=(0,2)$ and the $\CN=0$ and $\CN=(0,1)$ WZW models mentioned earlier, is that the target $G_\C$, which is topologically $T^* G$ or simply $G\times \mathfrak{g}$, is non-compact.
The NLSM into $G_\C$ most certainly lacks a normalizable vacuum, which complicates the analysis and invalidates the usual unitarity bounds (such as positivity of the level).
In noncompact bosonic models, the normalizable ground state often exists simply because the quantum corrections (such as zero point fluctuations of the modes normal to the classical moduli space) lift the flat directions.
This usually does not happen in SUSY models, and the question of existence of the normalizable vacuum may be quite hard (see, e.g., an old unsolved problem discussed in \cite{Sethi:1997pa,Moore:1998et}).
In 2D models with $(0,2)$ SUSY, this question is closely related to global symmetries and their anomalies, as we will see now.

Recall that a conserved flavor symmetry current $(j_z, j_{\bar z})$ must be incorporated into some sort of current multiplet. It is fairly clear that $j_z$ must be the bottom component of a $(0,2)$ superfield $\CK_z$, while for $j_{\bar z}$ there can be more than one possibility. Bertolini, Melnikov and Plesser found a useful formulation of the flavor current multiplet (we call it the BMP multiplet) in \cite{BMP} that consists of two superfields $(\CK_z,\CI)$. The bottom component of $\CK_z$ is indeed $j_z$, while $j_{\bar z}$ appears as the top component of $\CI$. The superfields obey two relations:
\begin{align}
\bar\partial \CK_z + \partial [D_+,\bar{D}_+]\CI=0,\quad \bar{D}_+ (\CK_z + 2\partial \CI)=0,
\end{align}
where $D_+$ and $\bar{D}_+$ are the standard SUSY-covariant derivatives. The first relation here is just the conservation law, and the second one identifies the holomorphic current in the $\bar{Q}_+$-cohomology.

\emph{Proposition:} If the spectrum of local operators includes a BMP current multiplet $(\CK_z,\CI)$ for a flavor symmetry with negative anomaly coefficient, then there is no normalizable vacuum.

\emph{Proof:} By contradiction, assume that the IR CFT is compact (i.e., has the normalizable vacuum).
The flavor symmetry enhances to the affine Kac-Moody (AKM) in the IR, splitting into the separately conserved left and right-moving affine currents of levels $(k_L, k_R)$. Compactness and unitarity imply $k_L\geq0$ and $k_R\geq0$. The anomaly, which also coincides with the level of the chiral algebra in the $\bar{Q}_+$-cohomology, can be identified with $k_L- k_R$ \cite{WittenPertur:2005}. Since it is negative by assumption, we necessarily have $k_R>0$, thus the right-moving AKM current $j_{\bar z}$ is non-trivial, and so is the IR limit of the superfield $\CI$ where it resides. The bottom component of $\CI$ is then a nontrivial dimension-zero operator.
The latter is impossible in a compact CFT, leading to a contradiction.

That the flavor symmetry is described by the BMP multiplet is related to the absence of SUSY enhancement. Any current $(j_z, j_{\bar z})$ can be extended to the pair of $(0,2)$ superfields $(\CK_z, \bar{\CK}_{\bar z})$ whose bottom components are $j_z$ and $j_{\bar z}$ (such a formulation was used in \cite{DedushChiral:2015}).
In the theory with $\CN=(0,2)$ supersymmetry, one expects the higher components of $\CK_z$ to vanish at the CFT point, leaving only the left-moving current $j_z$ nontrivial.
One cannot say the same about the right-moving multiplet $\bar{\CK}_{\bar z}$, which necessarily contains the nonzero superpartners of $j_{\bar z}$.
In particular, its dimension-$\frac32$ component \emph{could} be a new conserved supercurrent. The fact that in the BMP multiplet $\bar{\CK}_{\bar z}=[D_+,\bar{D}_+]\CI$ shows that the dimension-$\frac32$ component is merely a derivative of some dimension-$\frac12$ operator, not a new supercurrent. The multiplet $\CI$ is also known to describe the $N=2$ AKM algebras \cite{SKM1,Delius:1990st}, which can only exist in a noncompact CFT, as we see now. 

If a compact model has a symmetry described by the BMP current multiplet, then the superfield $\CI$, having dimension zero, must vanish in the IR, and the symmetry current becomes purely left-moving. In fact, it seems likely that in models without SUSY enhancement (both compact and non-compact), all flavor symmetries fit into the BMP current multiplets. We do not know how to prove this statement, so instead we formulate the following \emph{Conjecture:} Every flavor current in an $\CN=(0,2)$ theory without SUSY enhancement fits into the BMP current multiplet $(\CK_z, \CI)$. If we further find a symmetry with negative anomaly coefficient, the proposition proved earlier implies that the model is non-compact. Our $G_\C$ sigma model always has a symmetry with negative anomaly \eqref{anom02}, so according to this rule, it has no normalizable vacuum.

\vspace{-0.3cm}
\subsection{(Dual?) Landau-Ginzburg description}
It turns out that the $G_\C$ sigma-models, at least for classical groups, admit the Landau-Ginzburg (LG) formulations.
It is the simplest for $G=SU(N)$, $G_\C=SL(N,\C)$.
Let $M_i{}^j$ be a $(0,2)$ chiral multiplet valued in complex $N\times N$ matrices ${\rm Mat}(N,\C)$, and $\Gamma$ be one Fermi multiplet.
An LG model with such fields and the $(0,2)$ superpotential
\begin{equation}
\CW = \mu\Gamma (\det M - 1),
\end{equation}
where $\mu$ is a mass parameter, clearly flows to the NLSM with the target determined by $\det M=1$, i.e., $SL(N,\C)$.
Such an LG model manifestly has $SU(N)\times SU(N)$ symmetry, under which $M$ is the bi-fundamental and $\Gamma$ is neutral.
This model has $k=0$, i.e., no B-field (WZ term) in the action.
An easy way to generate the B-field can be found from the anomaly matching.
Indeed, the WZ term contributes $(k,-k)$ to the $SU(N)\times SU(N)$ anomaly in the IR, which can be matched in the UV by adding Fermi multiplets $\Lambda_a{}^i$, $a=1,\ldots,k$, $i=1, \ldots,N$ (a bifundamental under $U(k)\times SU(N)$), and chiral multiplets $\Phi_i{}^a$ (a bifundamental under $SU(N)\times U(k)$). Here $U(k)$ is an auxiliary symmetry that should be fully non-anomalous to disappear without a trace in the IR. We modify the $(0,2)$ superpotential as follows:
\begin{equation}
\label{WZLG}
\CW = \mu\Gamma (\det M - 1) + \mu\Tr \Lambda M \Phi.
\end{equation}
The first term ensures that we still flow to the moduli space $\det M=1$, i.e., the group $SL(N,\C)$ emerges as the target. In particular, $M$ is a non-degenerate matrix there. The second term tells us that the multiplets $\Lambda$ and $\Phi$ become massive on the moduli space, with $\mu M$ playing the role of mass matrix. At low energies (or for large enough $\mu$) we can integrate $(\Lambda, \Phi)$ out. Their contribution to the anomaly is precisely $k$ for the left $SU(N)$ global symmetry and $-k$ for the right (and no mixed or pure anomalies for $U(k)$). Since the anomaly must be saturated by \emph{something}, integrating out $\Lambda$ and $\Phi$ must indeed generate the WZ$_k$ term (plus some effective metric and the irrelevant higher-derivative terms). This is known to be true, and an explicit computation, at least for $G=SU(2)$, can be found in the literature \cite{Abanov:1999qz}. We thus conclude that \eqref{WZLG} is the simplest modification of the LG model that accounts for the WZ term. We may call it the LG-WZW model.

Such a model clarifies some aspects of the theory and muddens the others, as we will see in a companion paper \cite{DL2}.
For example, we study the perturbative chiral algebra, which has the structure of the affine current algebras $\hat{\mathfrak{g}}_{k-h^\vee} \oplus \hat{\mathfrak{g}}_{-k-h^\vee}$ extended by the bi-modules. Such an answer emerges perturbatively both in the NLSM and the LG descriptions, and is also well-motivated from the 3D viewpoint. 

At the nonperturbative level, however, the three perspectives do not obviously align. In the $\CN=(0,2)$ NLSM description, the known nonperturbative effects come from compact holomorphic curves \cite{Dine:1986zy,Dine:1987bq,Beasley:2005iu}, which are absent in $G_\C$. Nonetheless, the vortices \eqref{vortex_op} might generate new nonperturbative corrections, leaving the question of non-perturbative physics of the non-compact NLSM open-ended. The LG model and the 3D gauge theory on the interval provide two alternative UV completions of the $G_\C$ NLSM. On the one hand, in the LG description, the common belief is that the perturbative chiral algebra does not receive any nonperturbative corrections \cite{Witten:1993jg,Melnikov:2009nh,DedushChiral:2015,BMP,Gholson:2018zrl}. On the other, the gauge theory is known to possess boundary monopoles that non-trivially correct the boundary VOA \cite{Costello:2020ndc}, and thus the interval VOA as well \cite{DL2} (such monopoles becomes vortices \eqref{vortex_op} in the 2D limit). It seems that the LG model and the 3D gauge theory on the interval provide UV definitions of the $G_\C$ NLSM with different spectra of defects, and hence different nonperturbative physics. The latter is a conjecture that will hopefully be addressed elsewhere.
 
 
It is straightforward to write the BMP current multiplets for the $G\times G$ global symmetry of the LG description \cite{BMP}. For the left $SU(N)$ symmetry:
\begin{equation}
\begin{split}
\hspace*{-0.2cm}\CK_i{}^j = \sum_a \Lambda_a{}^j \bar{\Lambda}^a{}_i &- \frac{1}{2} \sum_k (M_i{}^k \partial\bar{M}^j{}_k - \bar{M}^j{}_k \partial M_j{}^k),\\
&\CI_i{}^j = -\frac14 \sum_k M_i{}^k \bar{M}^j{}_k,
\end{split}
\end{equation}
and for the right $SU(N)$:
\begin{equation}
\begin{split}
\tilde{\CK}_i{}^j &= -\frac{1}{2} \sum_a (\Phi_i{}^a\partial\bar{\Phi}^j{}_a - \bar{\Phi}^j{}_a\partial\Phi_i{}^a)\cr &\ \ \,\,- \frac{1}{2} \sum_k (M_k{}^j \partial\bar{M}^k{}_i - \bar{M}^k{}_i \partial M_k{}^j),\cr
\tilde{\CI}_i{}^j &= -\frac14 \sum_k M_k{}^j \bar{M}^k{}_i - \frac14 \sum_a \Phi_i{}^a \bar{\Phi}^j{}_a.
\end{split}
\end{equation}
The bottom components of $\CI$ and $\tilde{\CI}$ are clearly non-zero in the IR (we do not even need the fact that the anomaly \eqref{anom02} has negative coefficients). Thus such dimension-zero operators live in the noncompact sector of the model.

For completeness, we describe LG models for other classical groups. When $G=SO(N)$, $G_\C= SO(N,\C)$, the multiplets are: (1) chiral $N\times N$ matrix $M_i{}^{j}$; (2) Fermi symmetric matrix $\Upsilon_{ij}$; (3) neutral Fermi $\Gamma$; (4) $k$ fundamental chirals $\Phi_i{}^a$, $i=1..N$, $a=1..k$; (5) $k$ antifundamental Fermi's $\Lambda_a{}^i$. The superpotential is
\begin{equation}
\hspace*{-0.2cm}\CW=\mu\left[\Tr \Upsilon(MM^T - 1) + \Gamma (\det M - 1) + \Tr \Lambda M \Phi\right].
\end{equation}
Note that the role of $\Gamma$ here is very minor: to select $SO(N,\C)$ as opposed to $O(N,\C)$. If we wish to work with the $O$ group, the Fermi multiplet $\Gamma$ and the corresponding term in $\CW$ should be dropped.

For $G=Sp(N)$, $G_\C=Sp(N,\C)$, one fixes the symplectic matrix $\omega$, such that $M\in G_\C$ obeys $M^T \omega M = \omega$.
The multiplets are (1) $2N\times 2N$ matrix $M$ of chirals; (2) antisymmetric $2N\times 2N$ Fermi matrix $\Omega$; (3) $k$ fundamental chirals $\Phi_i{}^a$, $i=1..2N$, $a=1..k$; (4) $k$ antifundamental Fermi's $\Lambda_a{}^i$, $i=1..2N$, $a=1..k$. The superpotential is
\begin{equation}
\CW = \mu \left[\Tr \Omega(M^T\omega M - \omega) + \Tr \Lambda M \Phi \right].
\end{equation}

In all these models, one can similarly write the BMP multiplets for the $G\times G$ symmetry and identify the dimension-zero operators in the noncompact sector.

\section{Conclusions}
In this work, using a novel computational trick, we found explicitly the two-derivative effective action for the 3d $  \mathcal{N} = 2 $ pure YMCS theory dimensionally reduced on an interval. This yields a non-compact 2D $\CN=(0,2)$ NLSM into the complefixied gauge group $G_\C$, which flows to a noncompact SCFT. We also discussed the relation between non-compactness and the flavor current multiplet in the theory. An alternative UV completion for the $G_\C$ NLSM by the LG model was constructed for classical groups $G= SU(N), SO(N)$ and $Sp(N)$. We stress that in this paper we work strictly in the semiclassical regime.

Along the way we point out and speculate on many physical aspects of this model. Being a non-compact sigma model in two dimensions, it is in general subtle and lacks various nice properties that come with the normalizable vacuum, but it is still a consistent QFT. It would be interesting to understand the nonperturbative physics of this model, as there are hints from 3D that it is nontrivial. This would, in particular, address the case of small WZ level $|k|<h^\vee$ that is likely to break SUSY.

In the companion paper \cite{DL2}, we are studying the chiral algebra of this model, and give a more expanded motivation connecting the problem to VOA$[M_4]$.

\begin{acknowledgements}
\textbf{Acknowledgements:} We benefited from the useful discussions and/or correspondence with: A.~Abanov, T.~Dimofte, D.~Gaiotto, Z.~Komargodski, I.~Melnikov, N.~Nekrasov, M.~Ro{\v{c}}ek.
M.~L. is grateful to A.~Venkata for multiple insightful conversations.
\end{acknowledgements}


\bibliographystyle{apsrev4-1}
\bibliography{int1}
\end{document}